\documentclass[10pt,conference]{IEEEtran}

\usepackage{verbatim,amsmath,enumerate,amssymb,hhline}
\usepackage[square,comma,numbers]{natbib} 
\usepackage{graphicx}
\usepackage{amsmath,amssymb,xcolor,verbatim}
\usepackage{tikz}
\tikzset { domaine/.style 2 args={domain=#1:#2} }

\def\norm#1{\left\| #1 \right\|}

\newtheorem{thm}{Theorem}[section]
\newtheorem{lemma}[thm]{Lemma}
\newtheorem{prop}[thm]{Proposition}

\newtheorem{ex}{Example}
\newtheorem{remark}[thm]{Remark}

\def\diag{\textrm{diag}}

\DeclareMathOperator*{\argmin}{argmin}

\providecommand{\abs}[1]{\ensuremath{\left\lvert #1 \right\rvert}}
\providecommand{\norm}[1]{\ensuremath{\left\Vert #1 \right\Vert}}

\providecommand{\floor}[1]{\ensuremath{\left\lfloor #1 \right\rfloor}}
\providecommand{\ceil}[1]{\ensuremath{\left\lceil #1 \right\rceil}}

\providecommand{\vv}[1]{\textquotedblleft #1\textquotedblright}

\newcommand{\Q}{\mathbb{Q}}
\newcommand{\Z}{\mathbb{Z}}
\newcommand{\C}{\mathbb{C}}
\newcommand{\R}{\mathbb{R}}

\newcommand{\Hh}{\mathbb{H}}
\renewcommand{\IEEEQED}{\IEEEQEDopen}

\DeclareMathOperator{\SL}{SL}
\DeclareMathOperator{\SU}{SU}
\DeclareMathOperator{\SO}{SO}

\providecommand{\abs}[1]{\ensuremath{\left\lvert #1 \right\rvert}}
\providecommand{\norm}[1]{\ensuremath{\left\Vert #1 \right\Vert}}

\providecommand{\vv}[1]{\textquotedblleft #1\textquotedblright}
\newcommand*{\dotleq}{\mathrel{\dot{\leq}}}

\newcommand{\D}{{\mathcal D}}

\begin{document}

\title{Towards a complete DMT classification \\ of division algebra codes}
\author{
\IEEEauthorblockN{Laura Luzzi}
\IEEEauthorblockA{Laboratoire ETIS\\
 CNRS - ENSEA - UCP \\
Cergy-Pontoise, France \\
laura.luzzi@ensea.fr}
\and
\IEEEauthorblockN{Roope Vehkalahti}
\IEEEauthorblockA{Department of  Mathematics and Statistics\\
University of Turku\\
Finland\\
 roiive@utu.fi}
\and
\IEEEauthorblockN{Alexander Gorodnik }
\IEEEauthorblockA{School of Mathematics\\
 University of Bristol \\
United Kingdom \\
a.gorodnik@bristol.ac.uk}
}

\maketitle

\begin{abstract}
This work aims at providing new bounds for the diversity multiplexing gain trade-off of a general class of division algebra based lattice codes.  

In the low multiplexing gain regime, some bounds were previously obtained from the high signal-to-noise ratio estimate of the union bound for the pairwise error probabilities. 
Here these results are extended to cover a larger range of multiplexing gains.  The improvement is achieved by 
using ergodic theory in Lie groups to estimate the behavior of the sum arising from the union bound.
  
In particular, the new bounds for lattice codes derived from $\Q$-central division algebras suggest that these codes can be divided into  two subclasses based on their Hasse-invariants at the infinite places. Algebras with ramification at the infinite place seem to provide better diversity-multiplexing gain tradeoff.

\end{abstract}

\section{Introduction}
In \cite{VLL2013} the authors  proved that the union bound can be used to analyze the diversity - multiplexing gain trade-off (DMT) of a large class of division algebra based lattice codes. 
This work was based on upper bounding the pairwise error probability (PEP) in the high signal-to-noise ratio (SNR) regime and then analyzing the behavior of the union bound by combining information on the zeta function and on the distribution of units of the division algebra.

The choice to focus on the high SNR approximation of the PEP allowed 
to analyze the behavior of the union bound using algebraic methods. However,  it  also implicitly restricted the analysis to be effective only for low multiplexing gain levels.

In this work we will use a more accurate expression for the pairwise error and extend the earlier DMT analysis  to cover a larger range of multiplexing gains.
  When we have enough receiving antennas, we can cover the whole multiplexing gain region. For fewer receive antennas, we have bounds up to a certain  multiplexing gain threshold.

As previously in \cite{VLL2013} the proofs rely heavily on the fact that the codes under analysis are coming from division algebras. This allows us to attack this otherwise quite impenetrable question using analytic methods from the ergodic theory of Lie groups \cite{Strong_Wavefront}.

This work confirms that from the DMT point of view all the division algebra codes with complex quadratic center have equal (and optimal) diversity multiplexing gain curve.
When the center of the algebra is $\Q$, our work suggests that division algebra based lattice codes can be divided to two subclasses with respect to their DMT. 
The difference between these two subclasses is whether the Hasse invariant at the infinite place is ramified or not. In particular, division algebras with ramification lead to a better DMT.

Besides giving a new lower bound (that we believe to be tight) for the DMT of a general family of division algebra based lattice codes, this work also sheds some light on the applicability and limitations of the union bound approach in Rayleigh fading channels. In \cite[Section 3D]{ZT}  the authors speculate that the union bound  cannot be used to
measure the DMT of a coding scheme accurately. Our work reveals that if we have good enough understanding of the spectrum of the pairwise error probabilities, and we have enough receive antennas, even a naive union bound analysis can be used to analyze the DMT of a space-time code.

\section{Notation and preliminaries}

\subsection{Central division algebras}\label{basic}

Let  $\mathcal{D}$ be a degree $n$  $F$-central division algebra where $F$ is either $\Q$ or a quadratic imaginary field.  Let $\Lambda$ be an \emph{order} in $\mathcal{D}$ and  $\psi_{reg}: \mathcal{D} \to M_n(\C)$ the left regular representation of the algebra $\D$.
When the center  $F$ is complex quadratic, $\psi_{reg}(\Lambda)$  is a $2n^2$-dimensional lattice and when $F=\Q$ it is $n^2$-dimensional. We are now interested in the diversity multiplexing gain trade-off of coding schemes based of the lattices $\psi_{reg}(\Lambda)$.  When $F$ is complex quadratic, we can attack the question directly. However, in the case where the center is $\Q$ we will instead consider lattices $A\psi_{reg}(\Lambda)A^{-1}$, where  $A$ is a certain  matrix in 
$M_n(\C)$. While the performance of  schemes  derived from $A\psi_{reg}(\Lambda)A^{-1}$ and $\psi_{reg}(\Lambda)$  can be very different, the diversity-multiplexing gain curves are the same.

Consider matrices
$$
\begin{pmatrix}
A &   -B^* \\
B&  A^*                                               
\end{pmatrix}
\in M_{2n}(\C),
$$
where $*$ refers to complex conjugation and $A$ and $B$ are complex matrices in $M_n(\C)$. We denote this set of matrices by $M_n(\Hh)$.

We say that the algebra $\D$ is \emph{ramified at the infinite place} if 
$$
\D\otimes_{\Q}\R\simeq M_{n/2}(\mathbb{H}).
$$
If it is not, then
$$
\D\otimes_{\Q}\R\simeq M_{n}(\R).
$$

\begin{lemma} \cite[Lemma 9.10]{VLL2013}\label{embeddings}

If the infinite prime is ramified in the algebra $\D$, then there exist a matrix $A\in M_n(\C)$ such that
$$
A\psi_{reg}(\Lambda)A^{-1}\subset M_{n/2}(\mathbb{H}).
$$
If $\D$ is not ramified at the infinite place, then there exist a  matrix $B\in M_n(\C)$ such that
$$
B\psi_{reg}(\Lambda)B^{-1}\subset M_{n}(\R).
$$
\end{lemma}

From now on we will simply use notation $\psi$ for both embeddings of Lemma \ref{embeddings}, when the center is $\Q$ and  for $\psi_{reg}$, when the center is complex quadratic.

\subsection{System Model}

We consider a multiple-input multiple output (MIMO) system with $n$ transmit antennas and $m$ receive antennas, and minimal delay $T=n$. The received signal is given by
$$ Y=\sqrt{\frac{\rho}{n}} H \bar{X} + W,$$
where $\bar{X} \in M_n(\C)$ is the transmitted codeword, $H, W \in M_{m,n}(\C)$ are respectively the channel matrix and additive noise, both with i.i.d. circularly symmetric complex Gaussian entries $h_{ij}, w_{ij} \sim \mathcal{N}_{\C}(0,1)$, and $\rho$ is the signal-to-noise ratio. \\
In the DMT setting, we consider code sequences $\mathcal{C}(\rho)$ whose size grows with the signal-to-noise ratio. More precisely, the multiplexing gain $r$ is defined as 
$$r=\lim_{\rho \to \infty} \frac{1}{n}\frac{\log \abs{\mathcal{C}}}{\log \rho}.$$
Let $P_e$ denote the average error probability of the code. Then the diversity gain is given by 
$$d(r)=-\lim_{\rho \to \infty}\frac{\log P_e}{\log \rho}.$$

Let now $\Lambda$ be an order in  a degree $n$ $F$-central division algebra  $\D$ and $\psi$ an embedding as defined in Section \ref{basic}.

Given $M$, we consider the finite subset of elements with Frobenius norm bounded by $M$:
$$\Lambda(M)=\{ x \in \Lambda \;:\; \norm{\psi(x)} \leq M\}.$$
Let $k \leq 2n^2$ be the dimension of $\Lambda$ as a $\Z$-module. As in \cite{VLL2013}, we choose $M=\rho^{\frac{rn}{k}}$ and consider codes of the form $\mathcal{C}(\rho)=M^{-1} \psi(\Lambda(M))=\rho^{-\frac{rn}{k}} \psi(\Lambda(\rho^{\frac{rn}{k}}))$. The multiplexing gain of this code sequence is indeed $r$, and it satisfies the average power constraint
$$ \frac{1}{\abs{\mathcal{C}}} \frac{1}{n^2} \sum_{X \in \mathcal{C}} \norm{X}^2 \leq 1$$  
We suppose that the channel matrix $H$ is perfectly known at the receiver but not at the transmitter, and consider maximum likelihood decoding
$$\hat{X}=\argmin_{X \in \mathcal{C}} \norm{Y-HX}^2.$$
The error probability is the average over $H$ of the error probability for fixed $H$:
$$P_e(H)=\int_{M_{m,n}(\C)} P_e(H) p(H) d\lambda(H),$$
where $\lambda$ is the Lebesgue measure, and the density of $H$ is the product of Gaussian densities:
$$p(H)=\frac{1}{\pi^{m n}}\prod_{i=1}^{m} \prod_{j=1}^n e^{-\abs{h_{ij}}^2}$$
For fixed $H$, the union bound for the error probability gives
$$P_e(H)=\mathbb{P}\{\hat{X} \neq \bar{X} | H\} \leq \sum_{X \in \mathcal{C}, X \neq \bar{X}} \mathbb{P}\{ \bar{X} \to X |H\}.$$
The pairwise error probability is upper bounded by the Chernoff bound on the $Q$-function \cite{TSC}:
\begin{align*}
&\mathbb{P} \{ \bar{X} \to X |H\}  \leq e^{-\frac{\rho}{8n}\norm{H(\bar{X}-X)}^2}
\end{align*}
By linearity of the code, 
$$P_e(H) \leq \sum_{X \in M^{-1}\psi(\Lambda(2M)) \setminus \{0\}} e^{-\frac{\rho}{8n}\norm{HX}^2}.$$
Note that we can replace $\frac{\rho}{8n}$ by $\rho$ without affecting the DMT; the coefficient \vv{2} in the sum also does not affect the DMT and so 
$$ P_e(H) \dotleq \sum_{\substack{X \in \mathcal{C},\\ X \neq 0}} e^{-\rho\norm{HX}^2}=\sum_{\substack{X \in \psi(\Lambda(M)),\\ X \neq 0}} e^{-\rho^{1-\frac{2rn}{k}}\norm{HX}^2}.$$
By the dotted inequality we mean $f(\rho) \dotleq g(\rho)$ if 
$$
\lim_{\rho\to \infty}\frac{\log f(\rho)}{\log \rho} \leq \lim_{\rho\to \infty}\frac{\log g(\rho)}{\log \rho}.
$$
To simplify notation, we define $c=\rho^{1-\frac{2rn}{k}}$. 
\section{A new upper bound on the error probability}
We now consider a similar argument to our previous paper \cite{VLL2013}. Let $\mathcal{I}$ be a collection of elements in $\Lambda$, each generating a different right ideal, and let $\mathcal{I}(M)=\mathcal{I} \cap \Lambda(M)$. Thus, each nonzero element $x \in \Lambda(M)$ can be written as $x=zv$, with $v \in \Lambda^*$. Moreover, since by hypothesis the center $F$ of the algebra is $\Q$ or an imaginary quadratic field, we have that the  subgroup 
$$
\Lambda^1=\{ x\in \Lambda^*\;:\;  \det(\psi(x))=1\},
$$
of units of reduced norm $1$ in $\Lambda^*$ has finite index $j=[\Lambda^*:\Lambda^1]$ \cite[p. 211]{Kleinert}. Let $a_1, a_2, \ldots, a_j$ be coset leaders of $\Lambda^1$ in $\Lambda^*$.\\
We note that $\Gamma=\psi(\Lambda^1)$ is an arithmetic subgroup of a Lie group $G$. In our case $G$ is one of the groups $\SL_n(\C)$, $\SL_n(\R)$ or $\SL_{n/2}(\mathbb{H})$. \\
The previous sum can be rewritten as 
\begin{equation*} 
\sum_{x \in \mathcal{I}(M)} \sum_{i=1}^j \sum_{\substack{u \in \Gamma, \\ \norm{\psi(xa_i)u} \leq M}} e^{-c \norm{H \psi(xa_i)u}^2}.
\end{equation*}
Since $xa_i \in \Lambda$, we have $\abs{\det(\psi(xa_i))}=\abs{\det(\psi(x))} \geq 1$. For $i \in \{1,\ldots,j\}$, let's consider
$$g_i=\frac{\psi(xa_i)}{\det(\psi(xa_i))^{\frac{1}{n}}} \in G. $$
With a slight abuse of notation, $\forall a \in G$ we denote by $B_a(M)$ the \vv{shifted ball} in $G$:
$$B_a(M)=\{g \in G \;:\; \norm{ag} \leq M\}.$$  
Using the notation $d_x=\abs{\det(\psi(x))}^{\frac{1}{n}}$, we find
\begin{equation} \label{sum}
P_e(H) \dotleq \sum_{x \in \mathcal{I}(M)} \sum_{i=1}^j \sum_{\substack{u \in \Gamma, \\ u \in B_{g_i}(M/d_x)}} e^{-c d_x^2 \norm{H g_i u}^2},
\end{equation}
Using a simplified argument inspired by the Strong Wavefront Lemma in \cite{Strong_Wavefront}, we will now show that the sum (\ref{sum}) can be bounded by an integral over the corresponding ball in $G$. \\
Let $\mathcal{F}_{\Gamma}$ be the fundamental domain of $\Gamma$ in $G$, which is a compact polyhedron in $G$ containing the identity element $e$. Consequently, $R_{\Gamma}=\max_{g \in \mathcal{F}_{\Gamma}} \norm{g}$ is finite (and greater than $n=\norm{e}$). 
Suppose $g \in \mathcal{F}_{\Gamma}$. By submultiplicativity of the Frobenius norm, we have that $\forall a \in M_{m,n}(\C)$,
\begin{align*}
&\norm{ag} \leq \norm{a} \norm{g} \leq R_{\Gamma} \norm{a}. 
\end{align*}
In particular, we have that $\forall g \in \mathcal{F}_{\Gamma}$, $\forall x \in G$,
$$\sum_{\substack{u \in \Gamma, \\ u \in B_{x}(M)}} e^{-c \norm{a u}^2} \leq \sum_{\substack{u \in \Gamma, \\ u \in B_{x}(M)}} e^{-\frac{c}{R_{\Gamma}^2} \norm{a u g}^2}.$$
By integrating both sides over $\mathcal{F}_{\Gamma}$, we find
\begin{align*}
&\mu(\mathcal{F}_{\Gamma}) \sum_{\substack{u \in \Gamma, \\ u \in B_{x}(M)}}  e^{-c \norm{au}^2} \leq \sum_{\substack{u \in \Gamma, \\ u \in B_{x}(M)}} \int_{\mathcal{F}_{\Gamma}} e^{-\frac{c}{R_{\Gamma}^2} \norm{aug}^2} d\mu(g)=\\
&=\sum_{\substack{u \in \Gamma, \\ u \in B_{x}(M)}} \int_{u\mathcal{F}_{\Gamma}} e^{-\frac{c}{R_{\Gamma}^2} \norm{ag}^2} d\mu(g),
\end{align*}
where $\mu$ is the Haar measure over $G$. The last equality follows from the invariance of $\mu$ under $G$-action. \\
Note that the images $u\mathcal{F}_{\Gamma}$ are disjoint. 
If $g= ug'$ with $g' \in \mathcal{F}_{\Gamma}$ and $u \in B_{x}(M)$, 
\begin{align*}
\norm{xg}=\norm{xug'} \leq \norm{xu}\norm{g'} \leq M R_{\Gamma}
\end{align*}
 We have
$$\bigcup_{u \in B_x(M)} u\mathcal{F}_{\Gamma} \subset B_x(M R_{\Gamma}),$$
where the union is disjoint. We can conclude that 
$$ \sum_{\substack{u \in \Gamma, \\ u \in B_{x}(M)}}  e^{-c \norm{au}^2} \leq \frac{1}{\mu(\mathcal{F}_{\Gamma})} \int_{B_x(R_{\Gamma}M)} e^{-\frac{c}{R_{\Gamma}^2} \norm{ag}^2} d\mu(g).$$
Let $M_x=\frac{R_{\Gamma}M}{d_x}$. From (\ref{sum}), the error probability is upper bounded by 
\begin{align*}
&\int_{M_{m,n}(\C)} \frac{1}{\mu(\mathcal{F}_{\Gamma})} \sum_{x \in \mathcal{I}(M)} \sum_{i=1}^j \int_{B_{g_i}(M_x)} e^{-\frac{cd_x^2}{R_{\Gamma}^2} \norm{Hg_i g}^2} d\mu\, p(H) d\lambda\\
&=\frac{j}{\mu(\mathcal{F}_{\Gamma})} \sum_{x \in \mathcal{I}(M)} \int_{M_{m,n}(\C)}\int_{B(M_x)}  e^{-\frac{cd_x^2}{R_{\Gamma}^2}  \norm{Hg}^2} d\mu\, p(H) d\lambda
\end{align*}
Since the integrand is a measurable and non-negative function, by Tonelli's theorem we can exchange the two integrals. From the determinant bound in \cite{TSC}, we have that $\forall X \in M_n(\C)$,
$$\int_{M_{m,n}(\C)} e^{-c\norm{HX}^2} p(H) d\lambda(H)= \frac{1}{(\det(I+c X X^*))^{m}}.$$
Thus the error probability is bounded by
\begin{align*}
&\frac{j}{\mu(\mathcal{F}_{\Gamma})} \sum_{x \in \mathcal{I}(M)} \int_{B(M_x)} \int_{M_{m,n}(\C)}  e^{-\frac{cd_x^2}{R_{\Gamma}^2} \norm{Hg}^2} p(H) d\lambda d\mu(g)= \notag\\
&=\frac{j}{\mu(\mathcal{F}_{\Gamma})} \sum_{x \in \mathcal{I}(\rho^{\frac{rn}{k}})} \displaystyle\int_{B(M_x)} \frac{1}{\left(\det\Big(I+\frac{d_x^2}{R_{\Gamma}^2}\rho^{1-\frac{2rn}{k}}  gg^*\Big)\right)^{m}} d\mu 
\end{align*}
Our problem is now reduced to finding an asymptotic upper bound for the integral
\begin{align}
I_x=\displaystyle\int_G \frac{1}{\big(\det\big(I+\delta_x^2\rho^{1-\frac{2rn}{k}}  gg^*\big)\big)^{m}} \chi_{B\big(\frac{\rho^{\frac{rn}{k}}}{\delta_x}\big)}(g) d\mu(g) \label{I} 
\end{align}
where we have defined $\delta_x=\frac{d_x}{R_{\Gamma}}$ to simplify notation. Note that 
\begin{align}
P_e \leq \frac{j}{\mu(\mathcal{F}_{\Gamma})} \sum_{x \in \mathcal{I}(\rho^{\frac{rn}{k}})} I_x
\label{cosets}
\end{align}
 
In the cases we're interested in, $G$ is a connected noncompact semisimple Lie group with finite center and admits a Cartan decomposition $G=KA^+K$, where $K$ is a maximal compact subgroup of $G$, and $A^+=\exp(\mathfrak{a}^+)$, with $\mathfrak{a}^+$ the positive Weyl chamber associated to a set of positive restricted roots $\bar{\Phi}^+$. Given a root $\alpha \in \bar{\Phi}^+$, we denote its multiplicity by $m_{\alpha}$. The highest weight is the sum of positive restricted roots with their multiplicities: $\beta=\sum_{\alpha \in \bar{\Phi}^+} m_{\alpha} \alpha$.\\
The following identity holds for any function $f \in L^1(G)$ \cite{Gorodnik_Oh}: 
$$ \int_{G} f d\mu= \int_{K \times \mathfrak{a}^+ \times K} f(k \exp(a) k') \prod_{\alpha \in \bar{\Phi}^+} (\sinh\alpha(a))^{m_{\alpha}} dk da dk',$$
where $da$ and $dk$  are the Haar measures on $\mathfrak{a}^+$ and $K$ respectively.\\
Note that in (\ref{I}), the integrand $f$ is invariant by $K$-action both on the left and on the right since it only depends on the singular values of $g$. So by definition of the normalized Haar measure, 
$$ \int_{G} f d\mu= \int_{\mathfrak{a}^+} f(\exp(a)) \prod_{\alpha \in \bar{\Phi}^+} (\sinh\alpha(a))^{m_{\alpha}} da.$$  
The dominant term (as a function of $\rho$) of the integral (\ref{I}) corresponds to the highest term of the sum 
$$\prod_{\alpha \in \bar{\Phi}^+} (\sinh \alpha(a))^{m_{\alpha}} =\sum_{\xi} h_{\xi} e^{\xi(a)}$$
The highest term corresponds to $\xi=\beta$ \cite{Gorodnik_Oh}. Therefore the dominant term of the expression is
\begin{equation} \label{dominant_term}
\int_{G} f(\exp(a)) e^{\beta(a)} da.
\end{equation}

\section{DMT bounds for division-algebra based codes}
In this section we will prove the following DMT bounds for the three classes of codes introduced earlier. 

\begin{prop}{\emph{Case $F=\Q(\sqrt{-d})$, $G=\SL_n(\C)$.}} \label{prop_SLnC} Let $d^*(r)$ be the piecewise linear function taking values $[(n-r)(m-r)]^+$ when $r$ is a positive integer, with equation 
\begin{equation} \label{d_star}
d^*(r)=-(m+n-2\floor{r}-1)r+mn-\floor{r}(\floor{r}+1).
\end{equation}
The diversity-multiplexing gain trade-off for space-time codes arising from $2n^2$-dimensional division algebras with imaginary quadratic center $F=\Q(\sqrt{-d})$ is $d^*(r)$ provided that $m \geq 2 \ceil{r}-1$.
\end{prop}

The DMT $d^*(r)$ is optimal for space-time codes \cite{ZT}, and Proposition \ref{prop_SLnC} is well-known \cite{EKPKL}, but an alternative proof is included here for the sake of completeness.

\begin{prop}{\emph{Case $F=\Q$, $G=\SL_n(\R)$.}} \label{prop_SLnR}
Let $d_1(r)$ be the line segment connecting the points $(r,[(m-r)(n-2r)]^+)$ where $2r \in \Z$, with equation  
\begin{equation} \label{d1r}
d_1(r)=(-n-2m+2\floor{2r}+1)r+mn-\frac{\floor{2r}}{2}(\floor{2r}+1).
\end{equation}
The diversity-multiplexing gain trade-off for space-time codes arising from $k=n^2$-dimensional division algebras with center $\Q$ not ramified at the infinite place is 
$d_1(r)$ provided that $m \geq \ceil{2r}-\frac{1}{2}$.
\end{prop}

\begin{prop}{\emph{Case $F=\Q$, $G=\SL_{n/2}(\Hh)$.}} \label{prop_SLnH}
Suppose that $n$ is even. Let $d_2(r)$ be the piecewise linear function connecting the points $(r,[(n-2r)(m-r)]^+)$ for $r \in \Z$. 
The diversity-multiplexing gain trade-off for space-time codes from $n^2$-dimensional division algebras with center $\Q$ which are ramified at the infinite place is $d_2(r)$
provided that $m \geq 2\ceil{r}-1$. 
\end{prop}

\begin{remark}
The results in Propositions \ref{prop_SLnR} and \ref{prop_SLnH} are new. Although this proof only provides a lower bound, we conjecture that $d_1(r)$ and $d_2(r)$ are actually the DMTs for these space-time codes for all values of $r$.
\end{remark}

\begin{figure}
\begin{tikzpicture}[scale=3, yscale=0.5]
\draw[->] (0,0) -- (1.1,0);
\draw (1.1,0) node[right] {$r$};
\draw [->] (0,0) -- (0,2.2);
\draw (0,0) node[below] {$0$};
\draw (0.5,0) node[below] {$\frac{1}{2}$};
\draw (1,0) node[below] {$1$};
\draw (0,0.5) node[left] {$\frac{1}{2}$};
\draw (0,2) node[left] {$2$};
\draw (0,2.2) node[above] {$d(r)$};
\draw [thick, samples=200, domain=0:1] plot(\x,{(-3+2*floor(2*\x))*\x+2-floor(2*\x)*(floor(2*\x)+1)/2});
\draw [dashed, thick, samples=200, domain=0:1] plot(\x,{2*((-1)*(1-2*floor(\x))*\x+1-floor(\x)*(floor(\x)+1))});
\draw [dotted] (0,0.5) -- (0.5,0.5);
\draw [dotted] (0.5,0) -- (0.5,0.5);
\end{tikzpicture}
\caption{DMT lower bounds for $n^2$-dimensional lattices from division algebras over $\Q$ when $n=2$ and $m=1$ (solid line: unramified at the infinite place; dashed line: ramified at the infinite place).}  
\end{figure}
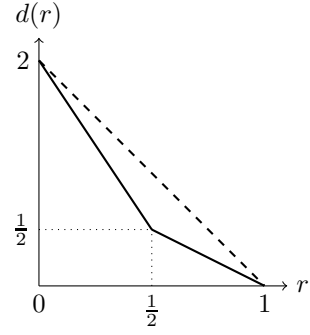

Before proceeding with the proofs, we need to give some details on the Lie group structures associated to the three main types of codes considered in this paper. See Appendix A in \cite{VLL2013} for definitions and details.
\begin{ex}{\emph{Case of center $F=\Q(\sqrt{-d})$, $G=\SL_n(\C)$.}}
The set of positive restricted roots is $\bar{\Phi}^+=\{e_i-e_k\}_{i<k}$, with multiplicity $m_{\alpha}=2$ for all $\alpha \in \bar{\Phi}^+$. 
Consider the algebra  $$\mathfrak{a}=\left\{ a=\diag(a_1,\ldots,a_n) \; : \; \sum\nolimits_{i=1}^n a_i=0\right\}.$$ 
The positive Weyl chamber associated to $\bar{\Phi}^+$ is
$$\mathfrak{a}^+=\left\{ a \in \mathfrak{a} \;:\; a_1 \geq a_2 \geq \cdots \geq a_n\right\}.$$
We have the Cartan decomposition $\SL_n(\C)=K \times A^+ \times K$, where $K=\SU_n$ and $A^+=\exp(\mathfrak{a}^+)$.\\
The 
highest weight is $\beta(a)=\sum_{i=1}^{n-1} 4(n-i)a_i$.  
\end{ex}

\begin{ex}{\emph{Case of center $F=\Q$, $G=\SL_n(\R)$.}}\\
We have $\bar{\Phi}^+=\{e_i-e_k\}_{i<k}$, with multiplicity $m_{\alpha}=1$ for all $\alpha \in \bar{\Phi}^+$. 
The positive Weyl chamber associated to $\bar{\Phi}^+$ is again
$\mathfrak{a}^+=\left\{ a \in \mathfrak{a} \; : \; a_1 \geq a_2 \geq \cdots \geq a_n\right\}$, and $\beta(a)=\sum_{i=1}^{n-1} 2(n-i)a_i$.
We have the Cartan decomposition $\SL_n(\R)=K \times A^+ \times K$, where $K=\SO_n$ and $A^+=\exp(\mathfrak{a}^+)$.
\end{ex}

\begin{ex}{\emph{Case of center $F=\Q$, $G=\SL_{n/2}(\Hh)$.}}\\
We suppose that $n=2p$ is even. 
Consider the algebra  
$\mathfrak{a}=\left\{ a=\diag(a_1,\ldots,a_p,a_1,\ldots,a_p) \; : \; \sum_{i=1}^p a_i=0\right\}.$ 
The set of positive restricted roots is $\bar{\Phi}^+=\{e_i-e_k\}_{1 \leq i<k <p}$, with multiplicity $m_{\alpha}=4$ for all $\alpha \in \bar{\Phi}^+$. The 
highest weight is $\beta(a)=8\sum_{i=1}^{p-1} (p-i)a_i$. The positive Weyl chamber associated to $\bar{\Phi}^+$ is
$\mathfrak{a}^+=\left\{ a \in \mathfrak{a} \; : \; a_1 \geq a_2 \geq \cdots \geq a_p\right\}.$
\end{ex}
Note that in all three cases, $\mathfrak{a}^+$ is a set of diagonal $n \times n$ matrices.

\begin{IEEEproof}[Proof of Propositions \ref{prop_SLnC}, \ref{prop_SLnR}, \ref{prop_SLnH}]
For the integral (\ref{I}), the dominant term (\ref{dominant_term}) is given by
\begin{align*}
&\int_{\mathfrak{a}^+} \frac{e^{\beta(a)}}{\prod_{i=1}^n (1+\delta_x^2\rho^{1-\frac{2rn}{k}} e^{2a_i})^m} \chi_{\big\{\sum\limits_{i=1}^n e^{2a_i} \leq \frac{\rho^{\frac{2rn}{k}}}{\delta_x^2}\big\}} da_1 \cdots da_{n-1} \\
& \leq \int_{\mathfrak{a}^+} \frac{e^{\beta(a)}}{\prod\limits_{i=1}^n (1+\delta_x^{2}\rho_x^{1-\frac{2rn}{k}} e^{2a_i})^m} \chi_{\big\{a_1 \leq \log\frac{\rho^{rn/k}}{\delta_x}\big\}} da_1 \cdots da_{n-1} 
\end{align*}
Note that the integral is only in $n-1$ variables and $a_n$ is just a dummy variable since $a_1 + a_2 + \cdots + a_n=0$. \\
Now consider the change of variables $a_i=b_i \log \left(\frac{\rho^{rn/k}}{\delta_x}\right)$. 
Given that $\delta_x \geq 1/R_{\Gamma}$, this integral is bounded by
\begin{equation*}
\left(\frac{rn}{k}\log \rho R_{\Gamma}\right)^{n-1} \int_{\mathcal{B}} \frac{e^{\beta(b)\log\frac{\rho^{rn/k}}{\delta_x}}}{\prod_{i=1}^n \big(1+e^{2(b_i-1)\log\frac{\rho^{rn/k}}{\delta_x}+\log \rho}\big)^m} db 
\end{equation*}
where 
$\mathcal{B}=\left\{ b \in \mathfrak{a}^+:\;  b_1 \leq 1\right\}.$\\
For our purposes, we can neglect logarithmic factors of $\rho$ in the sequel. \\
Let $(x)^+=\max(0,x)$. From the inequality $(1+e^x)^{-1} \leq e^{-(x)^+}$, we find the upper bound 
\begin{align*}
& \int_{\mathcal{B}} e^{\left[\beta(b) \log\frac{\rho^{rn/k}}{\delta_x} -m\sum\limits_{i=1}^n\left(2 (b_i-1)\log\frac{\rho^{rn/k}}{\delta_x}  +\log\rho\right)^{+}\right]} db=\\
&=\int_{\mathcal{B}} e^{\log \rho \left[\left(\frac{rn}{k}-\frac{\log\delta_x}{\log\rho}\right) \beta(b) -m\sum\limits_{i=1}^n\big(2 (b_i-1) \left(\frac{rn}{k}-\frac{\log\delta_x}{\log\rho}\right) +1\big)^{+}\right]} db =\\
&=\int_{\mathcal{B}} e^{-\log \rho \left[-\frac{sn}{k} \beta(b) +m\sum\limits_{i=1}^n\left(2\frac{sn}{k}(b_i-1) + 1\right)^{+}\right]} db_1 \cdots db_{n-1} 
\end{align*}
where $\frac{sn}{k}=\frac{rn}{k}-\frac{\log\delta_x}{\log\rho} \leq \frac{rn}{k}$. Note that $\mathcal{B}$ is contained in an $(n-1)$-dimensional cube with Lebesgue measure $1$. So our integral can be upper bounded by 
\begin{align*}
& \rho^{-\min\limits_{b \in \mathcal{B}} \left[-\frac{sn}{k} \beta(b) +m\sum_{i=1}^n\left( 2\frac{sn}{k}(b_i-1) + 1\right)^{+}\right]}= \\ 
& =\rho^{-\min\limits_{\alpha \in \mathcal{P}} \left[-\frac{\beta(\alpha)}{2} +m\sum_{i=1}^n\left( \alpha_i + 1 -\frac{2sn}{k}\right)^{+}\right]}. 
\end{align*}
where 
$\mathcal{P}=\left\{ \frac{2sn}{k} \geq \alpha_1 \geq \alpha_2 \geq \cdots \geq \alpha_n,\; \sum_{i=1}^n \alpha_i=0\right\}$, and $\alpha_i=b_i \frac{2sn}{k}$, $i=1,\ldots,n$.\\
Thus, we need to find 
\begin{align} 
&\bar{d}(s)=\min_{\alpha \in \mathcal{P}} g(\alpha), \quad \text{where} \notag\\
& g(\alpha)=-\frac{\beta(\alpha)}{2} +m\sum_{i=1}^n\left( \alpha_i + 1 -\frac{2sn}{k}\right)^{+}. \label{g}
\end{align}

The proof of the following two Remarks is elementary but rather tedious and can be found in the Appendix. 

\begin{remark}{(\emph{Case $G=\SL_n(\C)$}).} \label{min_SLnC}
On $\mathfrak{a}^+$, $\beta(\alpha)=-\sum_{i=1}^n 4 i \alpha_i$. In this case
\begin{align*} 
&g(\alpha)=\sum_{i=1}^n \left(2i\alpha_i+m\left(\alpha_i+1-\frac{s}{n}\right)^+\right),\\
& \mathcal{P}=\left\{ \frac{s}{n} \geq \alpha_1 \geq \alpha_2 \geq \cdots \geq \alpha_n,\; \sum_{i=1}^n \alpha_i=0\right\}.
\end{align*}
If $m \geq 2(\ceil{s}-1)$, then $\min_{\alpha \in \mathcal{P}} g(\alpha)=d^*(s)$. 
\end{remark}

\begin{remark}{(\emph{Case $G=\SL_n(\R)$}).} \label{min_SLnR}
On $\mathfrak{a}^+$, $\beta(\alpha)=-\sum_{i=1}^n 2 i \alpha_i$. In this case we have 
\begin{align*}
&g(\alpha)=\sum_{i=1}^n \left(i\alpha_i+m\left(\alpha_i+1-\frac{2s}{n}\right)^+\right),\\
&\mathcal{P}=\left\{ \frac{2s}{n} \geq \alpha_1 \geq \alpha_2 \geq \cdots \geq \alpha_n,\; \sum_{i=1}^n \alpha_i=0\right\}.
\end{align*}
If $m\geq \ceil{2s}-1$, then $\min_{\alpha \in \mathcal{P}} g(\alpha)=d_1(s)$. 
\end{remark}

The following Remark is more immediate. 

\begin{remark}{(\emph{Case $G=\SL_{n/2}(\Hh)$}).} \label{min_SLnH}
Let $n=2p$. Recall that $\mathfrak{a}=\left\{ a=\diag(a_1,\ldots,a_p,a_1,\ldots,a_p) \; : \; \sum_{i=1}^p a_i=0\right\}$, and $\beta(\alpha)=-8\sum_{i=1}^p i \alpha_i$ on $\mathfrak{a}^+$. We have $g(\alpha)=2\sum_{i=1}^{p} (2i\alpha_i+m(\alpha_i+1-\frac{s}{p})^+)$, and $\mathcal{P}=\left\{ \frac{s}{p} \geq \alpha_1 \geq \alpha_2 \geq \cdots \geq \alpha_p,\; \sum_{i=1}^p \alpha_i=0\right\}$. Note that the polyhedron and the function $g(\alpha)$ are very similar to the ones in Remark \ref{min_SLnC}. With the same reasoning, we find that the diversity order $\bar{d}(s)$ is lower bounded by the piecewise linear function connecting the points $(s,2(p-s)(m-s))=(s,(n-2s)(m-s))$ for $s \in \Z$, provided that $m \geq 2(\ceil{s}-1)$. 
\end{remark}

We can conclude that (neglecting logarithmic factors) the dominant term in $\rho$ in (\ref{I}) is of the order $f(\delta_x)$, where 
$$f(t)=\rho^{-\bar{d}(s)}=\rho^{-\bar{d}\left(r-\frac{k}{n}\frac{\log t}{\log\rho}\right)}.$$
Consequently, the dominant term in the error probability bound (\ref{cosets}) is bounded by 
\begin{align*}
&\frac{j}{\mu(\mathcal{F}_{\Gamma})} C(\log \rho R_{\Gamma})^{n-1} \sum_{x \in \mathcal{I}(\rho^{\frac{rn}{k}})} \rho^{-\bar{d}\left(r-\frac{k}{n}\frac{\log\delta_x}{\log\rho}\right)} 
\end{align*}
where $C$ is a constant independent of $\rho$ and $x$.\\
Recall that $\mathcal{I}$ is a collection of elements $x \in \Lambda$ generating distinct right ideals $x\Lambda$. We have
$$  \sum_{x \in \mathcal{I}(\rho^{\frac{rn}{k}})} f(\delta_x)=\sum_{x \in \mathcal{I}:\; \norm{\psi(x)}\leq \rho^{\frac{rn}{k}}} f(\delta_x) \leq \sum_{x \in \mathcal{I}:\; d_x\leq \rho^{\frac{rn}{k}}} f(\delta_x) $$
since by the arithmetic-geometric mean inequality, $d_x=\abs{\det(\psi(x))}^{\frac{1}{n}} \leq \norm{\psi(x)}$. Given $l \in \mathbb{N}$, define $s_l=\abs{\{x \in \mathcal{I} : l \leq  \delta_x < l+1\}}$, and $\forall t>0$, let $S_t=\sum_{l \leq t} s_l$. 
Since $f$ is decreasing and $\delta_x=d_x/R_{\Gamma} \leq d_x$, 
$$\sum_{x \in \mathcal{I}(\rho^{\frac{rn}{k}})} f(\delta_x) \leq \sum_{l \leq \rho^{\frac{rn}{k}}} s_l f(l).$$
Using summation by parts \cite[Theorem 1]{Tenenbaum}, we have
\begin{equation} \label{summation_by_parts}
\sum_{l \leq \rho^{\frac{rn}{k}}} s_l f(l)=S(\rho^{\frac{rn}{k}})f(\rho^{\frac{rn}{k}})-\int_{1}^{\rho^{\frac{rn}{k}}} S(t) f'(t) dt.
\end{equation} 
It is possible to show \cite[Theorem 29]{Gorodnik_Paulin} that given a central simple algebra $\mathcal{D}$ over $\Q$ and an order $\Lambda$ in $\mathcal{D}$, there exist constants $c,\delta>0$ such that 
$$\abs{\{ x \in \mathcal{I}: \;\; 1 \leq \abs{\det(\psi(x))} \leq A\}} = c A^{n}(1+O(A^{-\delta})).$$
Similarly, for a central simple algebra $\mathcal{D}$ over an imaginary quadratic field $F$ and an order $\Lambda$ in $\mathcal{D}$, $\exists c,\delta>0$ such that 
$$\abs{\{ x \in \mathcal{I}: \;\; 1 \leq \abs{\det(\psi(x))} \leq A\}} = c A^{2n}(1+O(A^{-\delta})).$$
In both cases, the exponent of $A$ is equal to $k/n$. Thus, in both cases we have $$S(t)= \abs{\{ x \in \mathcal{I}: \;\; 1 \leq \abs{\det(\psi(x))} \leq R_{\Gamma}^n t^n\}}  \sim t^{k}.$$  
Since $f(\rho^{\frac{rn}{k}})=\rho^{-\bar{d}(0)}=\rho^{-mn}$, the first term in (\ref{summation_by_parts}) is of the order $S(\rho^{\frac{rn}{k}})f(\rho^{\frac{rn}{k}}) \sim \rho^{-n(m-r)}$, which is smaller than $\rho^{-\bar{d}(r)}$ in the three cases we are considering.\\ Let's now focus on the second term in (\ref{summation_by_parts}), which can be written as 
\begin{align*}
&-\int_{1}^{\rho^{\frac{rn}{k}}} t^{k} \rho^{-\bar{d}\left(r-\frac{k}{n}\frac{\log t}{\log \rho}\right)} (\bar{d})'\left(r-\frac{k}{n}\frac{\log t}{\log \rho}\right) \frac{k}{nt} dt\\
&=-\log \rho \int_{0}^{r} \rho^{n(r-v)} \rho^{-\bar{d}(v)}(\bar{d})'(v) dv \leq \\
&\leq C\log \rho \int_0^{r} \rho^{nr-(nv+\bar{d}(v))} dv.
\end{align*}
after the change of variables $v=r-\frac{k}{n}\frac{\log t}{\log \rho}$, and recalling that $(\bar{d})'(v)\leq 0$. Define
$$d^{**}(v)=nv+\bar{d}(v).$$
To conclude the proof, we now deal with the three cases separately.

\paragraph{Case $G=\SL_n(\C)$}
$d^{**}(v)=nv+d^*(v)$ is a piecewise linear function interpolating the points of the parabola $v^2-mv+mn$ for $v \in \Z, v \leq \min(m,n)$. It is decreasing in $[0,v]$ provided that $d^{**}(\ceil{v}-1) \geq d^{**}(\ceil{v})$, or equivalently if the midpoint $\ceil{v}-\frac{1}{2} \leq \frac{m}{2}$.   \\
Assume that $m \geq 2\ceil{r}-1$. Then, we have 
$$\int_0^r \rho^{rn-d^{**}(v)}dv \leq r \rho^{rn-d^{**}(v)}=r\rho^{-d^*(r)},$$
and so $P_e(\rho) \dotleq \rho^{-d^*(r)}$.
\paragraph{Case $G=\SL_n(\R)$}
$d^{**}(v)=nv+d_1(v)$ is a piecewise linear function interpolating the points of the parabola $v^2-2mv+mn$ for $2v \in \Z, v \leq \min(m,\frac{n}{2})$. It is decreasing in $[0,v]$ provided that $d^{**}(\frac{\ceil{2v}}{2}-\frac{1}{2}) \geq d^{**}(\frac{\ceil{2v}}{2})$, or equivalently if the midpoint $\frac{\ceil{2v}}{2}-\frac{1}{4} \leq \frac{m}{2}$.   \\
Assume that $m \geq \ceil{2r}-\frac{1}{2}$. With the same reasoning as in the previous case we find
$P_e(\rho) \dotleq \rho^{-d_1(r)}$.
\paragraph{Case $G=\SL_{n/2}(\Hh)$}
$d^{**}(v)=nv+d_2(v)$ is a piecewise linear function interpolating the points of the parabola $2v^2-2mv+mn$ for $v \in \Z, v \leq \min(m,\frac{n}{2})$. It is decreasing in $[0,v]$ provided that $d^{**}(\ceil{v}-1) \geq d^{**}(\ceil{v})$, or equivalently if the midpoint $\ceil{v}-\frac{1}{2} \leq \frac{m}{2}$.   \\
Assume that $m \geq 2\ceil{r}-1$. Similarly to the previous cases we obtain  
$P_e(\rho) \dotleq \rho^{-d_2(r)}$.  
\end{IEEEproof}

\appendix
\subsection{Proof of Remark \ref{min_SLnC}} \label{Appendix_SLnC}

The function $g$ 
is a maximum of linear functions, and so it is piecewise linear and convex, but not necessarily concave. Note that $\mathcal{P}$ is an $(n-1)$-dimensional simplex bounded by the hyperplanes $
\bar{H}=\{\alpha_1 + \cdots + \alpha_n=0\}, \; H_0=\left\{\alpha_1=\frac{s}{n}\right\}, \; H_i=\{ \alpha_i=\alpha_{i+1}\},\; i=1,\ldots, n-1$.  
Each vertex of $\mathcal{P}$ is of the form 
$$V_k=\bar{H} \cap \left(\bigcap_{i \neq k} H_i\right), \quad k=0,\ldots, n-1.$$
We have $V_0=\mathbf{0}$, and $V_k$ is such that 
$$\alpha_1=\cdots=\alpha_k=\frac{s}{n}, \quad \alpha_{k+1}=\cdots=\alpha_n=-\frac{ks}{(n-k)n}.$$
Note that $g(\mathbf{0})=m(n-s)$, and 
\begin{align*}
g(V_k)=-ks +mk +m(n-k-s)^+.
\end{align*} 
If $s \in \Z$, $g(V_k) \geq g(V_{n-s})=(m-s)(n-s)=d^*(s)$. For non-integer $s$, we find that
\begin{align*}
&k<n-s \;\Rightarrow\; g(V_k) \geq g(V_{\floor{n-s}})=m(n-s)-s(\floor{n-s}),\\
&k>n-s\; \Rightarrow\;g(V_k) \geq g(V_{\ceil{n-s}})=(m-s)(\ceil{n-s}). 
\end{align*}
In both cases, $g(V_k)>d^*(s)$. \\
Since $g$ may not be concave, it may not a priori take its minimum on the vertices of $\mathcal{P}$. However, $g$ is piecewise linear on the subsets 
$$\mathcal{S}_k=\left\{\alpha \in \mathcal{P}: \quad \alpha_{k+1} \leq \frac{s}{n}-1, \; \alpha_k \geq \frac{s}{n} -1\right\}.$$
For $\alpha \in \mathcal{P}$, $\forall k \in \{1,\ldots,n\}$, we have
$$0=(\alpha_1+\ldots + \alpha_k) + (\alpha_{k+1} + \ldots + \alpha_n) \leq k\frac{s}{n} +(n-k) \alpha_{k+1},$$
which implies that 
\begin{equation} \label{condition}
\alpha_{k+1} \geq -\frac{sk}{n(n-k)} \quad \forall k \geq 1.
\end{equation}
Note that $\mathcal{S}_k$ has measure $0$ when $k \leq n-s$ because of the condition (\ref{condition}). So $\mathcal{P}=\bigcup_{k=n-s+1}^{n-1} \mathcal{S}_k$ and $$\min_{\mathcal{P}} g(\alpha)=\min_{n-s<k \leq n-1} \min_{\mathcal{S}_k} g(\alpha).$$
Since $g(\alpha)$ is linear on $\mathcal{S}_k$, its minimum in $\mathcal{S}_k$ is attained in one of the vertices. Therefore we need to check all the vertices of $\mathcal{S}_k$. The new vertices (that are not already vertices of $\mathcal{P}$) are the intersection of the hyperplane $\tilde{H}_k=\left\{\alpha_k=\frac{s}{n}
-1\right\}$ with the edges of $\mathcal{P}$. Let 
$$\mathcal{P}_k^+=\left\{ \alpha \in \mathcal{P} \;:\; \alpha_k \geq \frac{s}{n}-1\right\}, \quad \mathcal{P}_k^-=\mathcal{P}\setminus \mathcal{P}_k^+.$$ If there are $t$ vertices of $\mathcal{P}$ on one side of the hyperplane and $n-t$ vertices on the other side, the total number of new vertices is at most $t(n-t)$. 
For fixed $k>n-s$, we find that:
\begin{enumerate}
\item[-] $V_0 \in \mathcal{P}_k^+$; 
\item[-] for $j \geq k$, $V_j$ has $\alpha_k=\frac{s}{n}$ and so $V_j \in  \mathcal{P}_k^+$;
\item[-] if $j < n-s < k$, $V_j \in  \mathcal{P}_k^+$; 
\item[-] for $s \in \Z$, $V_{n-s} \in \tilde{H}_k$ so it's a vertex we've already checked; 
\item[-] for $n-s < j < k$, $V_j \in  \mathcal{P}_k^-$;
\end{enumerate}
Therefore the new vertices  $Q_{jl}$ and $R_{jl}$ arise from the edges  connecting $V_j$, $j \in \{\floor{n-s+1},\ldots, k-1\}$ with either $V_l$, $l \in \{0,\ldots, \ceil{n-s-1}\}$ or $V_l$, $l \in \{k, \ldots, n-1\}$, and these vertices are of the form
$$\tilde{H}_k \cap \bar{H} \cap \Big( \bigcap_{i \neq j,l} H_i\Big).$$ 
After some tedious calculations, we find that $Q_{jl}$ has coordinates
$\frac{s}{n}=\alpha_1=\ldots=\alpha_l> \alpha_{l+1} = \ldots = \frac{s}{n} + \frac{n-s-j}{j-l} = \alpha_j > \alpha_{j+1} = \ldots = \alpha_k = \ldots = \alpha_n=-1+\frac{s}{n}$
and $$g(Q_{jl})=m(n-s)-(n-j)(n-s)+l(n-s-j).$$ Recalling that $0 \leq l < n-s < j <k$, and letting $l=n-s-a$, $j=n-s+b$ with $a,b > 0$, we get 
$$g(Q_{jl})=(m-s)(n-s)+ab.$$
For $s \in \Z$, $g(Q_{jl})>(m-s)(n-s)=d^*(s)$. For $s \notin \Z$, the choice of $l$ and $j$ which minimizes $g(Q_{jl})$ is $\bar{l}=\floor{n-s}$, $\bar{j}=\ceil{n-s}$, and 
$$g(Q_{\bar{l}\bar{j}})=-s(m+n-2\floor{s}-1)-\floor{s}(\floor{s}+1)+mn=d^*(s).$$  
The points $R_{jl}$, where $n-s < j <k \leq l \leq n-1$, have coordinates
$ \frac{s}{n}=\alpha_1=\ldots=\alpha_j> \alpha_{j+1} = \ldots = \alpha_k= -1+ \frac{s}{n} = \ldots= \alpha_l > \alpha_{l+1} = \ldots = \alpha_n=\frac{l-j}{n-l}-\frac{ls}{n(n-l)}$
and $$g(R_{jl})= -sl+(l-j)(n-j)+mj.$$
Letting $j=n-s+a$, $l=j+b$, with $a>0$, $b \geq 1$, $a+b \leq s-1$ we find that 
$$g(R_{jl})=m(n-s)-s(n-s)+a(m-s-b).$$
We have $g(R_{jl})\geq d^*(s)$ provided that $m \geq 2(\ceil{s}-1)$. \hspace*{\fill}~\IEEEQED

\subsection{Proof of Remark \ref{min_SLnR}} \label{Appendix_SLnR}

The vertices of $\mathcal{P}$ are $V_0=\mathbf{0}$ and $V_k=(\alpha_1,\ldots,\alpha_n)$, $k=1,\ldots,n-1$, with 
$$\alpha_1=\cdots=\alpha_k=\frac{2s}{n}, \quad \alpha_{k+1}=\cdots=\alpha_n=-\frac{2ks}{(n-k)n}.$$
If $2s \in \Z$, then $g(V_k) \geq g(V_{n-2s})=(m-s)(n-2s)=d_1(s)$.\\
Suppose now that $2s \notin \Z$. For $k < n-2s$, 
$$g(V_k) \geq g(V_{\floor{n-2s}})= -(n-\ceil{2s})s+m(n-2s)\geq d_1(s),$$
with equality for $s \in (0,1/2)$.
For $k > n-2s$, we get 
$$g(V_k)\geq g(V_{\ceil{n-2s}})=(n-\floor{2s})(m-s) > d_1(s).$$
The function $g$ is piecewise linear on the subsets 
$$\mathcal{S}_k=\left\{\alpha \in \mathcal{P}: \quad \alpha_{k+1} \leq \frac{2s}{n}-1, \; \alpha_k \geq \frac{2s}{n} -1\right\},$$
that have positive measure for $k\geq n-2s$. The extra vertices of the region $\mathcal{S}_k$ (that are not vertices of $\mathcal{P}$) are the points $Q_{jl}$ and $R_{jl}$ connecting $V_j$, $n-2s<j<k \leq n$, with $V_l$, where $0\leq l <n-2s$ and $n-2s<j<k<l<n$ respectively. 
Note that since $n, j, k,$ and $l$ are integers, the points $Q_{jl}$ and $R_{jl}$ exist if and only if $\frac{1}{2} < s  \leq \frac{n}{2}$ and $\frac{3}{2} < s  \leq \frac{n}{2}$ respectively. \\
The point $Q_{jl}$ has coordinates
$\frac{2s}{n}=\alpha_1=\ldots=\alpha_l> \alpha_{l+1} = \ldots = \frac{2s}{n} + \frac{n-2s-j}{j-l} = \alpha_j > \alpha_{j+1} = \ldots = \alpha_k = \ldots = \alpha_n=-1+\frac{2s}{n},$ and
$g(Q_{jl})=m(n-2s)-(n-2s)\frac{(n-j)}{2}+\frac{l}{2}(n-j-2s)=(m-s)(n-2s)+\frac{(n-2s-l)(j-n+2s)}{2}$.\\
If $2s \in \Z$, note that $g(Q_{jl})>(m-s)(n-2s)$.\\ 
Suppose now that $2s \notin \Z$. Then $$g(Q_{jl}) \geq g(Q_{n-\floor{2s},n-\floor{2s}-1})=d_1(s).$$ 
Now let's consider the point $R_{jl}$, which has coordinates 
$ \frac{2s}{n}=\alpha_1=\ldots=\alpha_j> \alpha_{j+1} = \ldots = \alpha_k= -1+ \frac{2s}{n} = \ldots= \alpha_l > \alpha_{l+1} = \ldots = \alpha_n=\frac{l-j}{n-l}-\frac{2ls}{n(n-l)}$.
We have 
$$g(R_{jl})= -sl+(l-j)(n-j)+mj.$$
Letting $j=n-2s+a$, $l=j+b$, with $a > 0$, $b \geq 2$, $a+b \leq 2s-1$ we find that 
$$g(R_{jl})=m(n-s)-s(n-2s)+a(m-s-b/2).$$
If $2s \in \Z$, we have $b \leq 2s -2$ and $g(R_{jl}) \leq d_1(s)$ provided that $m \geq 2s-1$.\\ 
If $2s \notin \Z$, we have $b \leq \floor{2s}-1$ and $g(R_{jl}) \leq d_1(s)$ provided that $m \geq \floor{2s}=\ceil{2s}-1$. \hspace*{\fill}~\IEEEQED

\begin{small}

\end{small}


\begin{thebibliography}{10}
\bibitem{EKPKL} P. Elia, K. R. Kumar, P. V. Kumar, H.-F. Lu, and S. A. Pawar, ``Explicit Space-Time Codes Achieving the Diversity-Multiplexing Gain Tradeoff'', {\it IEEE Trans. Inf. Theory}, vol. 52, pp. 3869--3884, September 2006.
\bibitem{Gorodnik_Oh} A. Gorodnik, H. Oh, \vv{Orbits of discrete subgroups on a symmetric space and the Furstenberg boundary} \emph{Duke Math. J.} 139 (2007), no. 3, 483--525.
\bibitem{Strong_Wavefront} A. Gorodnik, H. Oh, N. Shah, \vv{Strong wavefront lemma and counting lattice points in sectors}, \emph{Israel J. Math.} 176 (2010), 419--444. 
\bibitem{Gorodnik_Paulin} A. Gorodnik, F. Paulin, \vv{Counting orbits of integral points in families of affine homogeneous varieties and diagonal flows}, \emph{Journal of Modern Dynamics} vol 8, n.1, pp 25--59, 2014.
\bibitem{Kleinert} E. Kleinert,  ``Units of classical orders: a survey'',  {\it L'Enseignement Math.} 40, pp. 205--248, 1994.
\bibitem{TSC} V. Tarokh, N. Seshadri, and A.R. Calderbank, ``Space-Time Codes for High Data Rate Wireless Communications:
Performance Criterion and Code Construction'', {\it IEEE Trans. Inf. Theory}, vol. 44, pp. 744--765, March 1998.
\bibitem{Tenenbaum} G. Tenenbaum, \emph{Introduction to analytic and probabilistic number theory}, Cambridge Studies in Advanced Mathematics, Cambridge University Press, 1995
\bibitem{VLL2013} R. Vehkalahti, H.-f. Lu, L. Luzzi,  ``Inverse Determinant Sums and Connections Between Fading Channel Information Theory and Algebra'', \emph{IEEE Trans. Inform. Theory}, vol 59,  pp. 6060--6082, September 2013.
\bibitem{ZT} L. Zheng and D. Tse, ``Diversity and Multiplexing: A Fundamental Tradeoff in Multiple-Antenna Channels'', {\it IEEE Trans. Inf. Theory}
vol. 49, pp. 1073--1096, May 2003.

\end{thebibliography}
\end{document}